\documentclass[twocolumn,nofootinbib]{revtex4}

\input psfig.sty
\usepackage[dvips]{color}
\usepackage{graphicx}
\usepackage{epsfig}

\begin{document}

\title{Evidence for cosmological particle creation?}

\author{C. Pigozzo$^{1,2}$, S. Carneiro$^{2}$, J. S. Alcaniz$^{3}$, H. A. Borges$^{2,4}$ and J. C. Fabris$^{5}$}

\affiliation{$^{1}$Imperial Centre for Inference and Cosmology, Imperial College, London, UK\\ 
$^{2}$Instituto de F\'{\i}sica, Universidade Federal da Bahia, Salvador, BA, Brasil\\ 
$^{3}$Departamento de Astronomia, Observat\'orio Nacional, Rio de Janeiro, RJ, Brasil\\ 
$^{4}$Institute of Cosmology and Gravitation, University of Portsmouth, Portsmouth, UK\\ 
$^{5}$Departamento de F\'{\i}sica, Universidade Federal do Esp\'{\i}rito Santo, Vit\'oria, ES, Brasil}

\begin{abstract}
A joint analysis of the linear matter power spectrum, distance measurements from type Ia supernovae and the position of the first peak in the anisotropy spectrum of the cosmic microwave background indicates a cosmological, late-time dark matter creation at $95\%$ confidence level.
\end{abstract}

\maketitle

In spite of its general agreement with current observations, the standard model based on conserved cold dark matter plus a cosmological constant ($\Lambda$CDM) presents some theoretical issues that motivate the study of more general models, either in the realm of modified gravity theories or in the context of General Relativity, by introducing extensions of the dark sector. Among those issues we can refer, for example, to the so-called ``coincidence problem", which is alleviated in models with interactions in the dark sector, models that have been deserving attention in the literature for a long time (see, for instance, \cite{Murgia} and references therein). Apart theoretical motivations, there are also observational reasons to include interactions in the dark sector, as the appearance of observational tensions when the standard model is tested against data of the large-scale clustering of galaxies (LSS), on one hand, and distance measurements of type Ia supernovae (SNe Ia), on the other. Some analyses of the linear mass power spectrum provide a value of $\Omega_{m0} \sim 0.2-0.3$ \cite{2dF,SDSS} for the present relative matter density, whereas from SNe Ia data one obtains $\Omega_{m0} \sim 0.3-0.4$ \cite{JLA,Trotta} (see also \cite{Planck} for an analysis of CMB data). With more recent galaxy surveys the value obtained for the matter density in the $\Lambda$CDM case presents a better agreement with that derived from distance measurements \cite{Hamman,Zhao}. Nevertheless, a tension still persists, particularly when LSS tests are compared to CMB results \cite{neutrinos}. 

This tension between different observational tests, if not related to systematic errors, may be interpreted as a signature of late-time dark matter creation. The idea is that, while the power spectrum depends strongly upon the matter density at the time of matter-radiation equality, which determines the spectrum turn-over and profile, distance measurements depend strongly upon the present value of the matter density. Therefore, if matter is created during the matter-dominated phase but the fitting model assumes matter conservation, the present matter density derived from the power spectrum will appear lower than that obtained from SNe Ia observations.
In order to verify this possibility, let us take, for simplicity, a substract formed only by pressureless dark matter with energy density $\rho$ and a vacuum term with energy density $\Lambda$ and equation of state $p_{\Lambda} = - \Lambda$. The balance equation for the total energy assumes the form
\begin{equation} \label{1}
\dot{\rho} + 3H\rho = - \dot{\Lambda} = \Gamma \rho,
\end{equation}
where $H = \dot{a}/a$ is the expansion rate and the second equality defines the rate of matter creation $\Gamma$. Clearly, a matter creation process is concomitant with a time decay of the vacuum term. Taking the derivative of the spatially flat Friedmann equation $\rho + \Lambda = 3H^2$ and substituting into (\ref{1}) we find $\dot{\Lambda} = 2 \Gamma \dot{H}$. In the particular case of a constant creation rate we obtain, apart from an integration constant, the linear decaying law $\Lambda = 2 \Gamma H$ \cite{tests}. 
In this {\emph{Letter}}, we consider a more general ansatz
\begin{equation} \label{2}
\Lambda = \sigma H^{-2\alpha},
\end{equation}
where $\alpha > -1$ and $\sigma = 3 (1 - \Omega_{m0}) H_0^{2(\alpha+1)}$. From the above equations, it is straightforward to show that $\Gamma = -\alpha \sigma H^{-(2\alpha+1)}$. Note that for $\alpha = -1/2$ the ansatz (\ref{2}) reduces to the linear case $\Lambda = 2 \Gamma H$ whereas for $\alpha = 0$ the standard model with constant $\Lambda$ and no matter creation is readily recovered. 
The background solution corresponding to the ansatz (\ref{2}) is given by~\cite{CP}
\begin{equation}\label{7}
\rho_T = 3H_0^2 \left[ 1-\Omega_{m0} + \frac{\Omega_{m0}}{a^{3(1+\alpha)}} \right]^{\frac{1}{1+\alpha}}.
\end{equation}
In the asymptotic future the cosmic fluid (\ref{7}) tends to a cosmological constant whereas for early times it behaves as pressureless matter, i.e.,
\begin{equation}\label{8}
\rho_T \approx \rho \approx 3H_0^2 \Omega_{m0}^{\frac{1}{1+\alpha}} z^3 \quad (z \gg 0)\;.
\end{equation} 
From this expression, it is clear that for the same density at high redshifts, negative values of $\alpha$ imply a higher present density as compared to the standard model, owing to the late-time matter creation process.

Before testing the hypothesis of cosmological matter creation against large-scale structure observations, it is necessary to verify whether the interacting vacuum clusters. For this purpose, let us rewrite Eq. (\ref{1}) in the covariant form
\begin{eqnarray}
T_{\,;\nu}^{\mu\nu} &=& Q^{\mu}, \label{3} \\
T_{\Lambda\,;\nu}^{\mu\nu} &=& - Q^{\mu}, \label{4} 
\end{eqnarray}
where $T$ and $T_{\Lambda}$ stands for the energy-momentum tensors of matter and vacuum, respectively. $Q^{\mu} = Q u^{\mu} + \bar{Q}^{\mu}$ is the energy-momentum transfer associated to the particle creation, decomposed here in components colinear and ortogonal to the cosmic fluid $4$-velocity $u^{\mu}$. Using $T_{\Lambda}^{\mu\nu} = \Lambda g^{\mu\nu}$, one obtains from (\ref{4})
\begin{eqnarray}
Q &=& - \Lambda_{,\nu} u^{\nu}, \label{5} \\
\bar{Q}^{\mu} &=& \Lambda_{,\nu} \left( u^{\mu} u^{\nu} - g^{\mu\nu} \right) \label{barQ}. \label{6}
\end{eqnarray}
In the fluid rest frame, the energy transfer is given by $Q = -\dot{\Lambda}$, as in (\ref{1}), whereas $\bar{Q}^{\mu} = 0$, which means that there is no momentum transfer in the isotropic background. A linear perturbation of equation (\ref{6}) leads to $\delta \bar{Q}^0 = 0$ and $\delta \bar{Q}_{i} = (\delta \Lambda + \dot{\Lambda} \theta)_{,i} = \delta \Lambda^c_{,i}$, where $\theta$ is the velocity potential ($\vec{\nabla}\theta = \delta \vec{u}$) and the second equality defines the gauge-invariant comoving perturbation of the vacuum component \cite{CB,Wands,Maartens}. Since the created dark matter is non-relativistic, the momentum transfer $\delta \bar{Q}_{i}$ is negligible compared to the energy transfer or, equivalently, $\delta \Lambda^c \approx 0$. In other words, this means that the vacuum energy does not cluster and all we observe in the large-scale clustering  is pressureless matter. This means, on the other hand, that there is no pressure term in the perturbed equations, which guarantees the absence of instabilities in the matter power spectrum. For a detailed analysis of the perturbed equations, see \cite{GCG} (for the particular case $\alpha = -1/2$ see also \cite{Velten}, where conserved baryons are explicitly included).

\begin{figure}
\includegraphics[trim= .1cm .5cm 0 3.5cm,height=7.5cm]{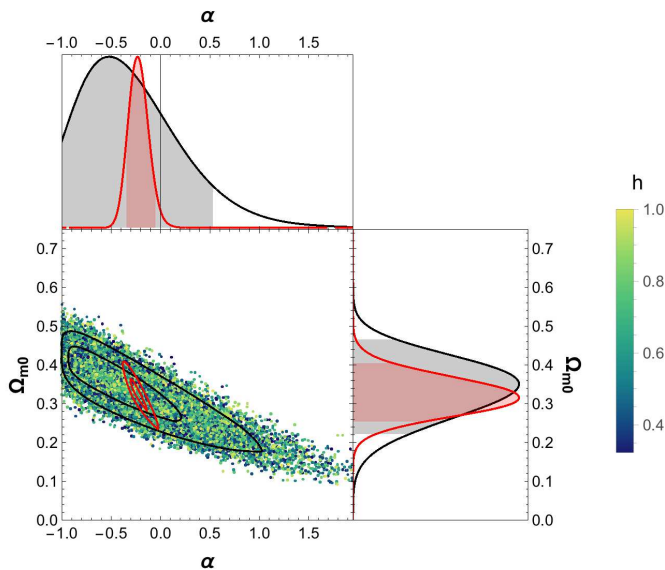}
\caption{
Probability distribution functions for $\alpha$ and $\Omega_{m0}$ from the joint analysis of SNe Ia + LSS + CMB (red). The black lines are PDFs for JLA supernovae only. The confidence regions in the $\alpha\times\Omega_{m0}$ plane are also shown.}
\label{fig1}
\end{figure}

Since there is no pressure term in the perturbed equations, all observed modes in the matter power spectrum evolve in the same way after the time of matter-radiation equality. Therefore, the present spectrum can be written as $P(k) = P_0 k^{n_s} T^2(k/k_{eq})$, where $k_{eq} \approx H(z_{eq})/z_{eq}$ is the comoving horizon scale at the time of matter-radiation equality, and $T(k/k_{eq})$ is a transfer function that gives the spectrum profile. Without loss of generality, we adopt $n_s = 1$ for the scalar spectral index and use the BBKS transfer function \cite{BBKS}. From (\ref{8}), the redshift of matter-radiation equality can be written as $z_{eq} = \Omega_{m0}^{{1}/{1+\alpha}}/\Omega_{R0}$, where $\Omega_{R0} = 4.15 \times 10^{-5} h^{-2}$ is the present relative density of radiation ($h = H_0/100$km/s-Mpc). We then obtain
\begin{equation} \label{10}
k_{eq} = 0.073\, Mpc^{-1}\, h^2 \Omega_{m0}^{\frac{1}{1+\alpha}}.
\end{equation}
The normalisation constant $P_0$ is a free parameter to be determined together with $\Omega_{m0}$, $h$ and $\alpha$.

We test our hypothesis of cosmological particle creation from a joint analysis of the linear matter power spectrum, the position of the first peak in the anisotropy spectrum of the cosmic microwave background, and the Hubble diagram of type Ia supernovae. For the matter power spectrum analysis we use data from the 2dFGRG survey \cite{2dF}. Compared to more recent surveys (see for example \cite{WiggleZ2}), the 2dFGRG data present here the advantage of being weakly dependent on the fiducial model used in their calibration. Updated surveys include data from lower scales and higher redshifts (up to $z \approx 1$) and then are more contaminated by the standard model, used in their calibration and in the treatment of non-linear effects. We also use the latest supernovae data, named Joint Light-Curve Analysis (JLA) compilation~\cite{JLA}, re-calibrated with the present model using the SALT2 fitter \cite{Salt2}.  The position of the first peak in the CMB spectrum is given by $l_1 = l_A (1 - \delta_1)$, where $l_A$ is the acoustic scale and $\delta_1 = 0.267 (r/0.3)^{0.1}$. In the latter expression, $r = \rho_R(z_{ls})/\rho(z_{ls})$ is the ratio between the radiation and matter densities at the redshift of last scattering $z_{ls} \approx 1090$ \cite{Hu}, written as
\begin{equation} \label{11}
r = \Omega_{R0} \Omega_{m0}^{-\frac{1}{1+\alpha}} z_{ls}.
\end{equation}
In our analysis we use $l_1 = 220.8 \pm 0.75$ \cite{WMAP}.
It is worth mentioning that the generalisations given by Eqs. (\ref{7}), (\ref{10}) and (\ref{11}) as well as the SNe Ia calibration mentioned earlier are essential for a proper analysis of the cosmological scenario here investigated. 

Using MultiNest \cite{multinest} we built chains with eight free parameters (four cosmological and four SNe Ia nuisanse parameters) and 5000 live points. The chains were analysed with GetDist\footnote{http://cosmologist.info/cosmomc.}. In Fig. \ref{fig1} we show the probability distribution functions for $\alpha$ and $\Omega_{m0}$ after marginalisation over the remaining parameters. 
Negative values of $\alpha$ are clearly favoured by the joint analysis, with the $2\sigma$ marginalised bounds on the model parameters 
given by $-0.39 < \alpha < -0.04$, $0.24 < \Omega_{m0} < 0.41$, and $0.66 < h < 0.74$. With the best-fit values for these parameters and the confidence interval for the normalisation constant $P_0$, we can also derive $\sigma_8$, the present rms fluctuation at a scale of $8h^{-1}$Mpc \cite{WL,GC}. Using a spherical top-hat filter \cite{Chani}, we obtain $0.81< \sigma_8 < 0.84$. Applying the same procedure to the $\Lambda$CDM case (i.e. by doing $\alpha = 0$), we obtain $\sigma_8 = 0.83 \pm 0.015$ ($2\sigma$), in agreement with Planck's result \cite{Planck}, which confirms the robustness of our method and surveys.

The standard $\Lambda$CDM model ($\alpha = 0$) is off by $> 95\%$ confidence level. This amounts to saying that, in spite of the simplicity of the standard phenomenology, more general models are favoured by these observations when the assumption of matter conservation is relaxed. This conclusion is corroborated by the JLA likelihood, which presents a maximum for $\alpha \approx -1/2$. Our results also agree with early joint analyses performed with other SNe Ia samples \cite{CP}. They should be confirmed with more recent galaxy surveys, involving a larger range of scales and redshifts. As mentioned above, this requires a full recalibration of data with the general model tested here, which is progress.

We are thankful to W. Zimdahl for discussions, and to CNPq, Faperj, Fapesb and Fapes for financial support.


\begin{thebibliography}{}

\bibitem{Murgia} R. Murgia, S. Gariazzo and N. Fornengo, JCAP {\bf 1604}, 014 (2016).

\bibitem[\protect\citeauthoryear{Cole et al.}{2005}]{2dF} S. Cole {\it et al.}, MNRAS {\bf 362}, 505 (2005).

\bibitem[\protect\citeauthoryear{Percival et al.}{2007}]{SDSS} W. J. Percival {\it et al.}, Astrophys. J. {\bf 657}, 645 (2007).

\bibitem[\protect\citeauthoryear{Betoule et al.}{2014}]{JLA} M. Betoule {\it et al.}, A\&A {\bf 568}, A22 (2014).

\bibitem[\protect\citeauthoryear{Sharif et al.}{2015}]{Trotta} H. Sharif, X. Jiao, R. Trotta and D. A. van Dyk, arXiv:1510.05954.

\bibitem[\protect\citeauthoryear{Ade et al.}{2015}]{Planck} P. A. R. Ade {\it et al.}, arXiv:1502.01589.

\bibitem{Hamman} J. Hamann {\it et al.}, JCAP {\bf 1007}, 022 (2010).

\bibitem{Zhao} G. B. Zhao {\it et al.}, MNRAS {\bf 436}, 2038 (2013).

\bibitem{neutrinos} R. A. Battye, T. Charnock and A. Moss, Phys. Rev. {\bf D91}, 103508 (2015).

\bibitem[\protect\citeauthoryear{Alcaniz et al.}{2012}]{tests} J. S. Alcaniz, H. A. Borges, S. Carneiro, J. C. Fabris, C. Pigozzo and W. Zimdahl, Phys. Lett. {\bf B716}, 165 (2012).

\bibitem[\protect\citeauthoryear{Carneiro \& Pigozzo}{2014}]{CP} S. Carneiro and C. Pigozzo, JCAP {\bf 1410}, 060 (2014).

\bibitem[\protect\citeauthoryear{Carneiro \& Borges}{2014}]{CB} S. Carneiro and H. A. Borges, JCAP {\bf 1406}, 010 (2014).

\bibitem[\protect\citeauthoryear{Wands et al.}{2012}]{Wands} D. Wands, J. De-Santiago and Y. Wang, Class. Quant. Grav. {\bf 29}, 145017 (2012).

\bibitem[\protect\citeauthoryear{Koyama et al.}{2009}]{Maartens}  K. Koyama, R. Maartens and Y. S. Song, JCAP {\bf 0910}, 017 (2009).

\bibitem{GCG} H. A. Borges, S. Carneiro, J. C. Fabris and W. Zimdahl, Phys. Lett. {\bf B727}, 37 (2013).

\bibitem{Velten} H. Velten, H. A. Borges, S. Carneiro, R. Fazolo and S. Gomes, MNRAS {\bf 452}, 2220 (2015).

\bibitem[\protect\citeauthoryear{Martin et al.}{2000}]{BBKS} J. Martin, A. Riazuelo and M. Sakellariadou, Phys. Rev. {\bf D61}, 083518 (2000).

\bibitem{WL} C. Heymans {\it et al.}, MNRAS {\bf 432}, 2433 (2013).

\bibitem{GC} P. A. R. Ade {\it et al.}, arXiv:1502.01597.

\bibitem{Chani} N. C. Devi, H. A. Borges, S. Carneiro and J. S. Alcaniz, MNRAS {\bf 448}, 37 (2015).

\bibitem[\protect\citeauthoryear{Parkinson et al.}{2012}]{WiggleZ2} D. Parkinson {\it et al.}, Phys. Rev. {\bf D86}, 103518 (2012).

\bibitem[\protect\citeauthoryear{Guy et al.}{2007}]{Salt2} J. Guy {\it et al.}, A\&A {\bf 466}, 11 (2007).

\bibitem[\protect\citeauthoryear{Hu et al.}{2001}]{Hu} W. Hu {\it et al.}, Astrophys. J. {\bf 549}, 669 (2001).

\bibitem[\protect\citeauthoryear{Spergel et al.}{2007}]{WMAP} D. N. Spergel {\it et al.}, Astrophys. J. Suppl. {\bf 170}, 377 (2007).

\bibitem[\protect\citeauthoryear{Feroz et al.}{2013}]{multinest} F. Feroz {\it et al.}, arXiv:1306.2144.

\end{thebibliography}
\end{document}